\documentclass[pre,twocolumn]{revtex4}

\usepackage{graphicx}
\usepackage{dcolumn}
\usepackage{bm}
\usepackage{amsmath}
\newcommand{\abin}{$\alpha$-BiN}

\begin{document}

\title{Modeling Discrete Combinatorial Systems as \\Alphabetic Bipartite Networks ($\alpha$-BiNs): Theory and 
Applications}

\author{Monojit Choudhury}
\affiliation{Microsoft Research India,196/36 2nd Main
Sadashivnagar, 560080 Bangalore, India.}

\author{Niloy Ganguly}%
\affiliation{Department of Computer Science and Engineering, Indian
Institute of Technology Kharagpur, 721302 Kharagpur, India.}

\author{Abyayananda Maiti}%
\affiliation{Department of Computer Science and Engineering, Indian
Institute of Technology Kharagpur, 721302 Kharagpur, India.}

\author{Animesh Mukherjee}%
\affiliation{Department of Computer Science and Engineering, Indian
Institute of Technology Kharagpur, 721302 Kharagpur, India.}

\author{Lutz Brusch}%
\affiliation{ZIH, TU Dresden, Zellescher Weg 12, 01069 Dresden, Germany.}%

\author{Andreas Deutsch}%
\affiliation{ZIH, TU Dresden, Zellescher Weg 12, 01069 Dresden, Germany.}%

\author{Fernando Peruani}
\email{fernando.peruani@iscpif.fr}
\affiliation{CEA-Service de Physique de l'Etat Condens\'{e}, Centre d'Etudes de Saclay, 91191 Gif-sur-Yvette, France, \\
Institut des Syst\'emes Complexes de Paris \^{I}le-de-France, 57/59, rue Lhomond, F-75005 Paris, France.}

\date{\today}

\begin{abstract}
Life and language are {\em discrete combinatorial systems} (DCSs) in
which the basic building blocks are finite sets of elementary units:
nucleotides or codons in a DNA sequence and letters or words in a language.
Different combinations of these finite units give rise to
potentially infinite numbers of genes or sentences. This type of
DCS can be represented as an Alphabetic Bipartite Network
($\alpha$-BiN)
where there are two kinds of nodes, one type represents the elementary units 
while the other type represents their combinations. There is an edge between a node
corresponding to an elementary unit $u$ and a node corresponding to
a particular combination $v$ if $u$ is present in $v$. Naturally, the partition
consisting of the nodes representing elementary units is fixed,
while the other partition is allowed to grow unboundedly.
Here, we extend recently  analytical findings 
for $\alpha$-BiNs derived in 
[Peruani et {\it al.}, Europhys. Lett. {\bf 79}, 28001 (2007)] 
 and empirically investigate two real world systems:
the codon-gene network and the phoneme-language network. The
evolution equations for $\alpha$-BiNs under different growth rules
are derived, and the corresponding degree distributions computed. 
It is shown that asymptotically the degree distribution of $\alpha$-BiNs can be described as a family of beta distributions.
The one-mode projections of the theoretical as well as the real
world $\alpha$-BiNs are also studied.
We propose a comparison of the real world degree distributions and 
our theoretical predictions as a means for inferring the mechanisms
underlying the growth of real world systems.
\end{abstract}

\maketitle


\section{\label{sec:intro} Introduction}

Two of the greatest wonders of evolution on earth, life and
language, are {\em discrete combinatorial systems}
(DCSs)~\cite{pinker}. The basic building blocks of DCSs are finite
sets of elementary units, such as the letters in language
and nucleotides (or codons) in DNA. Different combinations of these
finite elementary units give rise to a potentially infinite number of words
or genes.  
%
Here, we propose
a special class of complex networks as a model of DCSs. We shall refer to them as 
 {\em Alphabetic Bipartite Networks} (\abin s) in order to signify
the fact that the set of basic units, in both human and genetic
languages, can be considered as an {\em Alphabet}.

The \abin s  are a subclass of networks
where there are two different sets (partitions) of nodes: the bipartite networks. 
An edge, in a bipartite network, links nodes that appear in two different 
partitions, but never those in the same set. 
In most of the bipartite networks studied in the past both the 
partitions grow with time. Typical examples of this type of networks
include collaboration networks such as the
movie-actor~\cite{ramasco,watts_98,collaboration3,Alava:06,movie2},
article-author~\cite{collaboration4,collaboration6,lambiotte_05},
and board-director~\cite{caldarelli_05,strogatz_01} networks. In the
article-author network, for instance, the articles and authors are
the elements of the two partitions also known as the {\em ties} and
{\em actors} respectively. An edge between an author $a$ and an
article $m$ indicates that $a$ has co-authored $m$. The authors
$a$ and $a'$ are {\em collaborators} if both have coauthored the
same article, i.e., if both are connected to the same node $m$. The
concept of {\em collaboration} can be extended to represent, through bipartite networks, several diverse phenomena such as 
the city-people network~\cite{eubank_04}, in which an edge between a person and a
city indicates that the person has visited that particular city, the
word-sentence~\cite{cancho,Latapy}, bank-company~\cite{souma_03} or
donor-acceptor networks that account for injection and merging of
magnetic field lines~\cite{sneppen_04}.

Several models have been proposed to synthesize the structure of
these bipartite networks, i.e., when both the partitions grow
unboundedly over
time~\cite{ramasco,watts_98,collaboration3,Alava:06,Latapy}. It has
been found that for such growth models, when each incoming $tie$
node preferentially attaches itself to the $actor$ nodes, the
emergent degree distribution of the $actor$ nodes follows a
power-law~\cite{ramasco}. This result is reminiscent of unipartite
networks where preferential attachment results in power-law degree
distributions~\cite{barabasi}.

\begin{figure}
\centering \centering
\resizebox{\columnwidth}{!}{\rotatebox{0}{\includegraphics{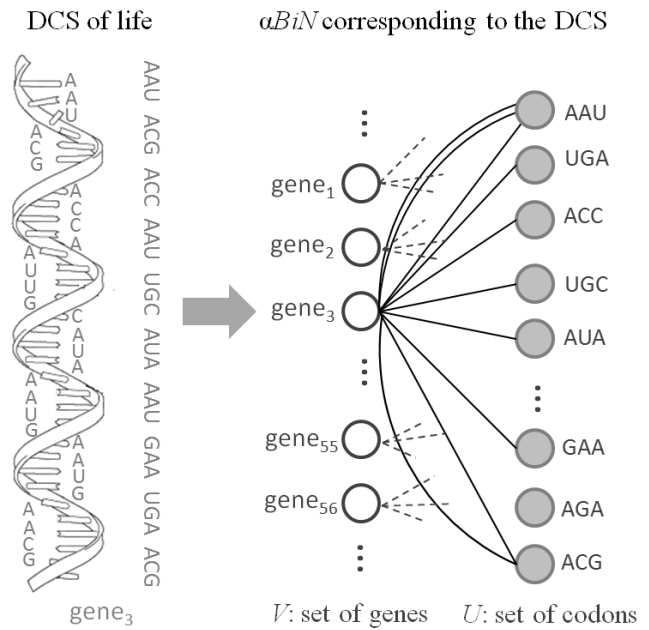}}}
\caption{DNA modeled as a bipartite network \abin . The
set $U$ consists of 64 codons, whereas the set $V$ of genes is
virtually infinite. Multiple occurrences of a codon in a gene have
been represented here by multi-edges. For instance, the codons `ACG'
and `AAU' have respectively 2 and 3 edges connecting to the node
gene$_3$. Alternatively, this could have been represented by single
edges with weights 2 and 3, while the weight of the other edges
would be equal to 1.} \label{fig:dcslife}
\end{figure}

On the other hand, bipartite networks where one of the partitions
remains fixed over time have received comparatively much less
attention. Since the set of basic units in DCSs is always finite and constant, \abin s have one of its partitions fixed, the one that represents the basic units (e.g. letters, codons). 
In contrast, the other partition, that represents the unique discrete combinations of basic unit (e.g., words, genes), can grow unboundedly over
time. 
Notice that the order in which the basic units are strung to form the
discrete combination is an important and indispensable aspect of the
system, which can be modeled within the framework of \abin s by
allowing ordering of the edges. Nevertheless, the scope of the
present work is limited to the analysis of unordered combinations.
Here we assume a word to be a bag of letters and a gene a multiset
of codons.
Fig.~\ref{fig:dcslife} illustrates the concepts through the example
of genes and codons.

A first systematic and analytical study of \abin s has been
presented in~\cite{us_epl}, where a growth model for such networks
based on preferential attachment coupled with a tunable randomness
component has been proposed and analyzed \footnote{Numerical evidence of the non-scale free character of the degree distribution of this type of system was first reported in~\cite{dahui_05}.}.
For sequential attachment,
i.e., when the edges are incorporated one by one, the exact
expression for the emergent degree distribution has been derived.
Nevertheless, for parallel attachment, i.e., when multiple edges are
incorporated in one time step, only an approximate expression has
been proposed. It has been shown that for both the cases, the degree
distribution approaches a beta-distribution asymptotically with
time. Depending on the value of  the randomness parameter four
distinct types of distributions can be observed; these, in
increasing order of preferentially, are: (a) normal distribution,
(b) skewed normal distribution with a single mode, (c) exponential
distribution, and (d) U-shaped distribution.

In this article, we briefly review these findings and extend the
analytical framework. We derive the exact growth model for parallel
attachment and study the degree distribution of the one-mode
projection of the network onto the alphabet nodes. These analytical
findings are further applied to study two well-known DCSs from the
domain of biology and language. We observe that in the codon-gene
network (codons are basic units or alphabet, genes are the discrete
combinations), the higher the complexity of an organism, the higher
the value of the randomness parameter. Similarly, the theory can
also satisfactorily explain the distribution of consonants over the
languages of the world studied through the phoneme-language network
(phonemes are the basic units and the sound systems of  languages
are the discrete combinations). Nevertheless, the study also reveals
certain limitations of the current growth models. For instance, we
observe that the topological characteristics of the network of
co-occurrence of phonemes, which is the one-mode projection of the
aforementioned network, is different from the theoretical
predictions.
This indicates that although the simple preferential attachment
based growth model succeeds in explaining the degree distribution of
the basic units of \abin , the theory fails to describe the one-mode
projection, which indicates that the real dynamics of the system is
much more complex.

The rest of the article is organized as follows: Sec.~\ref{sec:abin}
formally defines \abin\ and 
introduces two growth models and their corresponding theoretical
analysis. The two real networks -- codon-gene and phoneme-language
-- their topology and comparison with the theoretical model are
described in Sec.~\ref{sec:cogen} and \ref{sec:planet} respectively.
In Sec.~\ref{sec:disc} we summarize the obtained results, discuss the broader consequences of the
present work and propose some applications and alternative
perspectives on the same.

\section{Theoretical framework for \abin s}\label{sec:abin}
%
%


\subsection{Formal definition and modeling}
A bipartite graph $G$ is a 3-tuple $\langle U,V,E \rangle$, where
$U$ and $V$ are mutually exclusive finite sets of nodes (also known
as the two partitions) and $E \subseteq U\times V$ is the set of
edges that run between these partitions. We can also define $E$ as a
multiset whose elements are drawn from $U\times V$. Clearly, this
construction allows multiple edges between a pair of nodes and the
number of times the nodes $u \in U$ and $v \in V$ are connected can
be assumed to be the weight of the edge $(u,v)$. Note that although
we are defining $E$ to be a set of ordered tuples, the ordering is
an implicit outcome of the fact that edges only run between nodes in
$U$ and $V$. In essence, we do not mean any directedness of the
edges.

\abin s are a special type of bipartite networks where one of the partitions represents a set of basic units while the other partition represents their combinations. The set of
basic units is essentially finite and fixed over time. Let us denote
the unique basic units by the nodes in $U$. Let each
unique discrete combination of the basic units be denoted as a node
in $V$. There exists an edge between a basic unit $u \in U$ and a
discrete combination $v \in V$ iff $u$ is a part of $v$. If $u$
occurs in $v$ $w$ times, then there are $w$ edges between $u$ and
$v$, or alternatively, the weight of the edge $(u,v)$ is $w$.
Fig.~\ref{fig:dcslife} illustrates these concepts through the
example of genes and codons.

Notice that the above model overlooks the order in which the basic
units are strung into a particular discrete combination. The order
can be taken into account by labeling the basic units in order of
appearance in each element of $V$. However, in this work, we consider
only unordered versions of DCSs. As we shall see subsequently,
several real world DCSs, such as the phoneme-language network, are,
in fact, unordered sets.


\subsection{Growth model for sequential attachment}\label{sec:growthmodel}

In this subsection, we review the results derived in~\cite{us_epl}
which apply to sequential as well as parallel attachment. While the
results for sequential attachment are exact, for parallel attachment
they represent an approximation. In the next subsection the results
obtained in~\cite{us_epl} are extended and the exact derivation for
parallel attachment is presented.

The growth of \abin s is described in terms of a simple model based
on preferential attachment coupled with a tunable randomness
parameter. Suppose that the partition $U$ has $N$ nodes labeled as
$u_1$ to $u_N$. At each time step, a new node is introduced in the
set $V$ which connects to $\mu$ nodes in $U$ based on a predefined
attachment rule. Let $v_i$ be the node added to $V$ during the $i$th
time step. The theoretical analysis  assumes that $\mu$ is a
{\it constant} greater than 0. This constraint will be relaxed during
synthesis of the empirical networks. However, note that if the
degrees of the nodes in $V$ are sampled from a Poisson-like
distribution with mean $\mu$, the theoretical analysis holds good
asymptotically.

Let $\widetilde{A}(k_{i}^{t})$ be the probability of attaching a new
edge to a node $u_i$, where $k_{i}^{t}$ refers to the degree of the
node $u_i$ at time $t$. $\widetilde{A}(k_{i}^{t})$ defines the
attachment kernel that takes the form:
\begin{equation}\label{eq:kernel_attachment}
\widetilde{A}(k_{i}^{t}) = \frac{\gamma k_{i}^{t} + 1}{\sum_{j=1}^N
(\gamma k_{j}^{t} + 1)} \
\end{equation}
where the sum in the denominator runs over all the nodes in $U$, and
$\gamma$ is the tunable parameter which controls the relative weight
of preferential to random attachment. Thus, the higher the value of
$\gamma$, the lower the randomness in the system. Since in a
bipartite network the sum of the degrees of the nodes in the two
partitions are equal, the denominator in the above expression is
equal to $\mu\gamma t + N$. Note that the numerator of the
attachment kernel could be rewritten as $k_{i}^{t} + \alpha$, where
$\alpha=1/\gamma$ is a positive constant usually referred to as the
{\em initial attractiveness}~\cite{mendes}.

Physically this means that when a new discrete combination, say a
gene, enters the system, it is always assumed to have $\mu$ basic
units, e.g., a chain of $\mu$ codons. The patterns of the codons
constituting the newly entered gene depends on the prevalence of the
codons in the pre-existing genes as well as a randomness factor
$1/\gamma$. At this point it is worthwhile to distinguish between a
few basic sub-cases of the growth model. When $\mu = 1$, addition of
a node in $V$ is equivalent to addition of one edge in the network
and thus the edges attach to the nodes in $U$ in a sequential
manner. However, for $\mu > 1$ addition of an edge is no longer a
sequential process; rather $\mu$ edges are added simultaneously. We
refer to the former process as {\em sequential attachment} and the
latter as {\em parallel attachment}. Depending on the underlying
DCS, the parallel attachment process can be further classified into
two sub-cases. If it is required that the $\mu$ nodes chosen are all
distinct, then we call this {\em parallel attachment without
replacement}. On the other hand, if $v_i$ is allowed to attach to
the same node more than once, we refer to the process as {\em
parallel attachment with replacement} \footnote{The names {\it with} and {\it without replacement} refer to the fact that in the {\it without replacement} case, when a basic unit $u_k$ has been selected by one of the $\mu$ edges of node $v_i$, that basic unit is removed from the set of available basic units for the next edges of $v_i$. In contrast, in the {\it with replacement} case, if $u_k$ is selected, it is {\it replaced} back in the set of available basic units for the next edges of $v_i$. So, the same basic unit can be selected more than once by the same $V$ node.}. Thus, parallel attachment
without replacement leads to \abin s without multi-edges or weighted
edges, while parallel attachment with replacement results in \abin s
with multi-edges. The two cases collapse for the case of sequential
attachment.
To motivate the reader further, we provide some examples of natural
DCSs from each of the aforementioned classes.
\begin{itemize}
\item {\em Sequential attachment}: Since in the sequential attachment model, every node in $V$ has only one edge, it 
is not a discrete combination at all. Rather, each incoming $v_i$ is a reinstantiation of some basic unit $u_j$. 
However, think of a system where $U$ is the set of languages and $V$ is the set of speakers, and an edge between $u 
\in U$ and $v \in V$ implies that $u$ is the mother tongue of $v$. Although not a DCS, these type of ``class and its 
instance" systems are plentiful in nature and can be aptly modeled using sequential attachment.
\item {\em Parallel attachment with replacement}: Any DCS modeled as a sequence of the basic units can be thought to 
follow the ``with replacement" model. For instance, a gene can have many repetitions of the same codon and similarly, 
there may be multiple occurrences of the same word in a sentence.
\item {\em Parallel attachment without replacement}: A DCS that is a set of the basic units can be conceived as an 
outcome of the ``without replacement" model. For instance, the consonants and vowels (partition $U$) that form the 
repertoire of basic sounds (phonemes) of a language (partition $V$), proteins ($U$) forming protein complexes ($V$), 
etc.
\end{itemize}

In this work, we focus on the topological properties of \abin s that
are synthesized using the sequential and parallel attachment with
replacement. Nevertheless, in section \ref{sec:planet} we also
present some empirical results for the parallel attachment without
replacement model in the context of the phoneme-language network.

Any \abin\ has two characteristic degree distributions corresponding
to its two partitions $U$ and $V$. Here we assume that each node in
$V$ has degree $\mu$ and concentrate on the degree distribution of
the nodes in $U$. Let $p_{k,t}$ be the probability that a randomly
chosen node from  the partition $U$ has degree $k$ after $t$ time
steps. We assume that initially all the nodes in $U$ have a degree 0
and there are no nodes in $V$. Therefore,
\begin{equation}\label{eq:initial_condition}
p_{k,0} = \delta_{k,0}
\end{equation}

Here, $\delta$ represents the Kronecker symbol. It is interesting to
note that unlike the case of standard preferential attachment based
growth models for unipartite (e.g., the BA model~\cite{barabasi})
and bipartite networks (e.g.,~\cite{ramasco}), the degree
distribution of the partition $U$ in \abin s cannot be solved using
the stationary assumption that in the limit $t \to \infty$,
$p_{k,t+1} = p_{k,t}$. This is because the average degree of the
nodes in $U$, which is $\mu t/N$, diverges with $t$, and
consequently, the system does not have a stationary state.

In~\cite{us_epl} it has been shown that $p_{k,t}$ can be
approximated for $\mu \ll N$ and small values of $\gamma$ by
integrating:
\begin{equation}\label{eq:markovchain}
p_{k,t+1} = (1 - A_p(k,t)) p_{k,t} + A_p(k-1,t) p_{k-1,t} \
\end{equation}
where $A_p(k,t)$ is defined as
\begin{equation} \label{eq:kernel_mu_gt_1}
A_p(k,t)= \left\{
\begin{array}{ccc}
\frac{\left( \gamma k + 1 \right)\mu}{\gamma \mu t + N} & \mbox{for} & 0 \leq k \leq \mu t \\
0 & \mbox{otherwise} & \\
\end{array} \right.
\end{equation}
for $t>0$ while for $t=0$, $A_p(k,t)=(\mu/N) \delta_{k,0}$. The
numerator contains a $\mu$ because at each time step there are $\mu$
edges that are being incorporated into the network rather than a
single edge. The solution of Eq. (\ref{eq:markovchain}) with the
attachment kernel given by Eq. (\ref{eq:kernel_mu_gt_1})  reads:
\begin{equation}\label{eq:solutionparallel}
p_{k,t} = \left( \begin{array}{c} t \\ k \end{array}\right)
\frac{\prod_{i=0}^{k-1}{\left(\gamma i+ 1\right)}
\prod_{j=0}^{t-1-k}{\left( \frac{N}{\mu} - 1 + \gamma
j\right)}}{\prod_{m=0}^{t-1}{\left(\gamma m + \frac{N}{\mu}\right)}
} \
\end{equation}
As already mentioned in \cite{us_epl}, Eq. (\ref{eq:markovchain})
cannot describe the stochastic parallel attachment exactly because
it explicitly assumes that in one time step a node of degree $k$ can
only get converted to a node of degree $k+1$. Clearly, the
incorporation of $\mu$ edges in parallel allows the possibility for
a node of degree $k$ to get converted to a node of degree $k+ \mu$.
So, for $\mu>1$, Eq. (\ref{eq:solutionparallel}) is just an
approximation of the real process for $\mu \ll N$ and small values
of $\gamma$.
However, for $\mu=1$, i.e. for sequential attachment, Eq.
(\ref{eq:solutionparallel}) is the exact solution of the process.

Interestingly, for $\gamma > 0$, Eq. (\ref{eq:solutionparallel})
approaches, asymptotically with time, a beta-distribution as
follows.
\begin{equation}\label{beta}
p_{k,t} \simeq  C^{-1}\left( k/t \right)^{\gamma^{-1}-1} \left(1-
k/t \right)^{\eta-\gamma^{-1}-1} \
\end{equation}
Here, $C$ is the normalization constant. By making use of the
properties of beta distributions, we learn that depending on the
value of $\gamma$, $p_{k,t}$ can take one of the following
distinctive functional forms.

a) $\gamma=0$, a binomial distribution whose mode shifts with time,

b) $0<\gamma<1$, a skewed (normal) distribution which exhibits a
mode that shifts with time,

c) $1 \leq \gamma \leq (N/\mu) - 1$, a monotonically decreasing
(near exponential) distribution with the mode frozen at $k=0$, and

 d) $\gamma>(N/\mu) - 1$, a u-shaped
distribution with peaks at $k=0$ and $k=t$.

\begin{figure}
\centering\resizebox{\columnwidth}{!}{\rotatebox{0}{\includegraphics{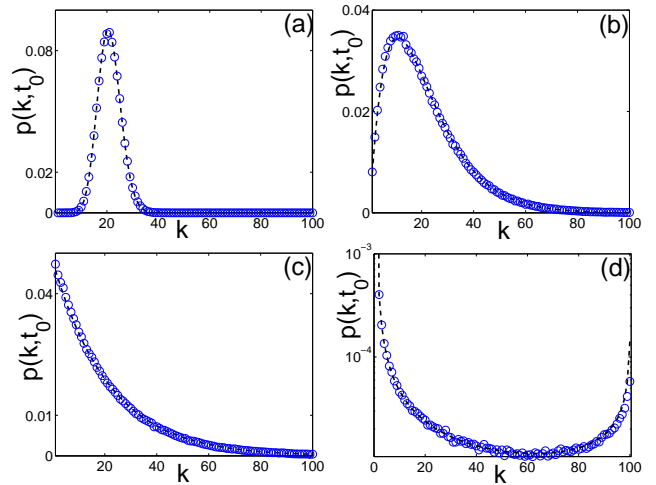}}}
\caption{The four possible degree distributions depending on
$\gamma$ for sequential attachment (and approximated expression for
parallel attachment).
Symbols represent average over $5000$, in (a)-(c), and $50000$, in
(d), stochastic simulations. The dashed curve is the theory given by
Eq. (\ref{eq:solutionparallel}).
From (a) to (c), $t_0=1000$, $N=1000$ and $\mu=20$.
(a) at $\gamma=0$, $p(k,t)$ becomes a binomial distribution.
(b) $\gamma=0.5$, the distribution exhibits a maximum which shifts
with time for $0\leq\gamma < 1$.
(c) $\gamma=1$, $p(k,t)$ does no longer exhibit a shifting maximum
and the distribution is a monotonically decreasing function of $k$
for $1\leq \gamma\leq (N/\mu) - 1$.
(d) $\gamma=2500$, $t_0=100$, $N=1000$ and $\mu=1$. $p(k,t)$ becomes
a u-shaped curve for $\gamma > (N/\mu) - 1$. }
\label{fig:transition_snapshots}
\end{figure}

\noindent Fig. \ref{fig:transition_snapshots} illustrates the
possible four regimes of Eq. (\ref{eq:solutionparallel}).
In the next subsection we present a generalization of
Eq.~(\ref{eq:markovchain}) for $\mu > 1$, i.e. for parallel
attachment, and solve it.

\subsection{Growth model for parallel attachment with replacement}
Recall that for parallel attachment, $t$ refers to the event of
introducing a new node in $V$ with $\mu$ edges. Therefore, the
correct expression for the evolution of $p_{k,t}$ has the form:
\begin{eqnarray}\label{eq:parallel}
p_{k,t+1} = (1 - \sum_{i=1}^{\mu}{\widehat{A}(k,i,t)}) p_{k,t} +
\sum_{i=1}^{\mu}{\widehat{A}(k-i,i,t)} p_{k-i,t} \
\end{eqnarray}
where $\widehat{A}(k,i,t)$ represents the probability at time $t$ of
a node of degree $k$ of receiving $i$ new edges in the next time
step.
The term $\sum_{i=1}^{\mu}{\widehat{A}(k,i,t)} p_{k,t}$ describes
the number of nodes of degree $k$ at time $t$ that change their
degree due to the attachment of $1$, $2$, $\ldots$, or $\mu$ edges.
On the other hand, nodes of degree $k$ will be formed at time $t+1$
by the nodes of degree $k-1$ at time $t$ that receive $1$ edge,
nodes of degree $k-2$ at time $t$ that receive $2$ edges, and so on.
This process is described by the term
$\sum_{i=1}^{\mu}{\widehat{A}(k-i,i,t)} p_{k-i,t}$.

\begin{figure}
\centering\resizebox{\columnwidth}{!}{\rotatebox{0}{\includegraphics{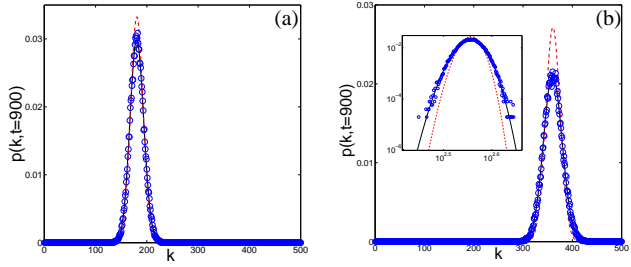}}}
\caption{Comparison for random attachment ($\gamma=0$) between the
approximation given by Eq. (\ref{eq:solutionparallel}) (dashed red
curve), the exact solution given by the integration of Eq.
(\ref{eq:mu_recursion}) (solid black curve), and stochastic
simulations. Symbols correspond to average over $500$ simulations.
In both the figures $N=100$. (a) corresponds to $\mu=20$ while (b)
to  $\mu=40$. The inset in (b) shows in log-log scale the deviation
of the approximation with respect to the exact solution and
simulations.} \label{fig:parallel_gamma0}
\end{figure}

\begin{figure}
\centering\resizebox{\columnwidth}{!}{\rotatebox{0}{\includegraphics{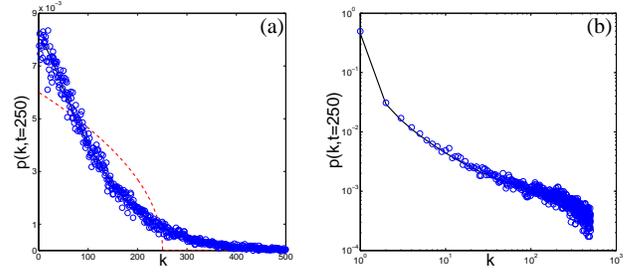}}}
\caption{Comparison for strong preferential attachment ($\gamma \geq
1$) between the approximation given by Eq.
(\ref{eq:solutionparallel}) (dashed red curve), the exact solution
given by the integration of Eq. (\ref{eq:mu_recursion}) (solid black
curve), and stochastic simulations (circles),averaged over 500 runs, for
parallel attachment for $\gamma \ge 1$. In both the figures $N=100$
and $\mu=40$. (a) corresponds to $\gamma=1$ while (b) to
$\gamma=16$. Notice that in (b) the approximation falls out of the
range of the figure, while the exact solution given by the
integration of Eq. (\ref{eq:mu_recursion}) describes the simulation
data quite well.} \label{fig:parallel_gamma_geq0}
\end{figure}
%

Next we derive an expression for $\widehat{A}(k,i,t)$. We start out
by a simple case: $\gamma=0$. Since in this case the probability for
an edge of attaching to a node is independent of its degree, if we
add $\mu$ edges, the probability for a node of receiving a single
edge is $\mu (1/N)(1 - 1/N)^{\mu-1}$, the probability of receiving
two edges is $\left( \begin{array}{c} \mu \\ 2 \end{array}\right)
(1/N)^{2} (1 - 1/N)^{\mu-2}$, and for the general case we obtain the
expression:
\begin{equation}\label{eq:kernel_real_parallel_RA}
\widehat{A}(k,i,t) = \left( \begin{array}{c} \mu \\ i
\end{array}\right) \left(\frac{1}{N}\right)^{i} \left(1- \frac{1}{N}
\right)^{\mu -i} \
\end{equation}
To extend this result to $\gamma>0$, we recall that if we add a
single edge, the probability for a node of degree $k$ of receiving
that edge is $\phi = \left( \gamma k +1 \right)/\left( \mu \gamma t
+ N \right)$, where we have assumed that previous to this edge we
had added $\mu t$ edges to the nodes in $U$. Clearly, $1-\phi$ is
the probability for the edge to attach to some other node. Taking
this into account, Eq.~(\ref{eq:kernel_real_parallel_RA}) is
generalized for $\gamma \geq 0$ as
\begin{equation}\label{eq:kernel_real_parallel}
\widehat{A}(k,i,t) = \left( \begin{array}{c} \mu \\ i
\end{array}\right) \left(\frac{\gamma k + 1}{\mu \gamma t +
N}\right)^{i} \left(1- \frac{\gamma k + 1}{\mu \gamma t + N}
\right)^{\mu -i}
\end{equation}
Inserting expression (\ref{eq:kernel_real_parallel}) into Eq.
(\ref{eq:parallel}), we obtain:
\begin{small}
\begin{eqnarray}\label{eq:parallel_full}
&& p_{k,t+1}  \\
&&\nonumber = \left[ 1 - \sum_{i=1}^{\mu}{\left( \begin{array}{c} \mu \\
i \end{array}\right) \left(\frac{\gamma k + 1}{\mu \gamma t + N}\right)^{i} \left(1- \frac{\gamma k + 1}{\mu \gamma t 
+ N} \right)^{\mu -i} } \right] p_{k,t} \\
&& \nonumber + \sum_{i=1}^{\mu}\left( \begin{array}{c} \mu \\ i
\end{array}\right) \left(\frac{\gamma \left( k-i \right) + 1}{\mu
\gamma t + N}\right)^{i}  \left(1- \frac{\gamma \left( k-i \right) +
1}{\mu \gamma t + N} \right)^{\mu -i}  p_{k-i,t}
\end{eqnarray}
\end{small}

The terms between parenthesis in the first line of Eq.
(\ref{eq:parallel_full}) can be simplified recalling that
\begin{eqnarray} \nonumber
&&\left(1- \frac{\gamma k + 1}{\mu \gamma t + N} \right)^\mu \\
&&\nonumber  = 1 - \sum_{i=1}^{\mu}{\left( \begin{array}{c} \mu \\ i
\end{array}\right) \left(\frac{\gamma k + 1}{\mu \gamma t +
N}\right)^{i} \left(1- \frac{\gamma k + 1}{\mu \gamma t + N}
\right)^{\mu -i} }
\end{eqnarray}

Therefore, Eq.~(\ref{eq:parallel_full}) can be rewritten by
including $i=0$ in the sum, whereby we obtain
\begin{eqnarray}\nonumber
p_{k,t+1} = \sum_{i=0}^{\mu}\left( \begin{array}{c} \mu \\ i
\end{array}\right) \left( \frac{\gamma \left( k-i
\right) + 1}{\mu \gamma t + N}\right)^{i} \\
\label{eq:mu_recursion} \left(1- \frac{\gamma \left( k-i \right) +
1}{\mu \gamma t + N} \right)^{\mu -i}  p_{k-i,t}
\end{eqnarray}
Note that Eq.~(\ref{eq:mu_recursion}) is a generalization of Eq. (2)
in~\cite{us_epl}, which can be obtained from
Eq.~(\ref{eq:mu_recursion}) by assigning $\mu =1$.

In Fig. \ref{fig:parallel_gamma0} we compare for random attachment
($\gamma=0$) the approximation given by Eq.
(\ref{eq:solutionparallel}) (dashed red curve), the exact solution
given by the integration of Eq. (\ref{eq:mu_recursion}) (solid black
curve), and stochastic simulations.

Notice that the approximation, as mentioned above, deviates from the
exact solution and simulations as $\mu$ increases. Looking at
Fig.~\ref{fig:parallel_gamma0} one may wrongly conclude that the
exact solution given by Eq. (\ref{eq:mu_recursion}) is a very minor
improvement over the approximation given by Eq.
(\ref{eq:solutionparallel}). However, for large values of $\gamma$
(see Fig. \ref{fig:parallel_gamma_geq0}), i.e., for
strong preferential attachment, Eq. (\ref{eq:solutionparallel})
drastically fails to describe the simulation data, while Eq.
(\ref{eq:mu_recursion}) accurately explains the data. To summarize, Figs. \ref{fig:parallel_gamma0} and
\ref{fig:parallel_gamma_geq0} validate Eq. (\ref{eq:mu_recursion})
and show that, except when $\mu \ll N$ and small $\gamma$, the
only way to describe the degree distribution for parallel attachment
with replacement is through the integration of Eq.
(\ref{eq:mu_recursion}). 
%


\subsection{One-mode projection}

In this section, we analyze the degree distribution of the one-mode
projection of \abin s onto the set $U$.
Formally, for an \abin\ $\langle U,V,E \rangle$, the one-mode
projection onto $U$ is a graph $G_U: \langle U, E_U\rangle$, where
$u_i, u_j \in U$ are connected (i.e., $(u_i,u_j) \in G_U$) if there
exists a node $v \in V$ such that $(u_i,v) \in E$ and $(u_j,v) \in
E$.
If there are $w$ such nodes in $V$ which are connected to both $u_i$
and $u_j$ in the \abin\  $G$, then there are $w$ edges linking $u_i$
and $u_j$ in the one-mode projection $G_U$. Alternatively, one can
conceive of a weighted version of $G_U$, where the weight of the
edge $(u_i,u_j)$ is $w$. In the context of the codon-gene network,
the one-mode projection is a codon-codon network, where two codons
are connected by as many edges as there are genes in which both of
these codons occur. The one-mode projection of an \abin\  provides
insight into the relationship between the basic units. 
For instance in linguistics the one-mode projection of the word-sentence \abin\  
reveals the co-occurrence of word pairs, which in turn 
provides crucial information about the syntactic and semantic 
properties of the words (see, for example, \cite{ferrer:04syntax,cancho}).

In \cite{arbit_dd:01} a general technique for computing the degree
distribution of the one-mode projection of a bipartite network is
described. The method has been derived by making use of the concept
of generating functions. As we shall see shortly, this technique is
only suitable for estimating the weighted degree distribution of the
one-mode projection. 

Here, we propose a novel technique to derive
the {\em thresholded degree distribution} of the one-mode projection
for any arbitrary threshold.
We start out by studying first the simple cases of the (non-thresholded) degree distribution of the one-mode projection for 
sequential and parallel attachnment, to finally focus on the new technique to derive the degree distribution of the thresholded one-mode projection for parallel attachment. 
Notice 
that in order to distinguish the degree distributions of the
one-mode projection from their bipartite counterpart, we shall use
the symbol $p_u(k,t)$ to refer to the probability that a randomly
chosen node from the one-mode projection of an \abin\ with $t$ nodes
in $V$ (i.e., after $t$ time steps) has degree $k$.

\subsubsection{Sequential attachment}
Recall that in the sequential attachment based growth model only one
edge is added per time step and consequently, every node in $V$ has
degree $\mu = 1$. Therefore, for any two nodes in $U$, say $u_i$ and
$u_j$, there is no node $v \in V$, which is connected to both $u_i$
and $u_j$ (this is because degree of $v$ is 1). Thus, for \abin s
that have been grown using the sequential attachment model, the
one-mode projection is a degenerate graph with $N$ nodes and 0
edges. The degree distribution of this network is
\begin{equation}\label{eq:unidd_seq}
p_u(k,t) = \delta_{k,0}
\end{equation}

\subsubsection{Parallel attachment}
Recall that in the parallel attachment model, at each time step the
node which is added to $V$ has $\mu$ edges. Consider a node $u \in
U$ that has degree $k$ in the \abin . Therefore, $u$ is connected to
$k$ nodes in $V$, each of which is connected to $\mu-1$ other nodes
in $U$. Defining the degree of a node as the number of edges
attached to it, in the one-mode projection, $u$ has a degree of
$q=k(\mu-1)$. Consequently, the degree distribution of $G_U$,
$p_u(q,t)$, is related to $p_{k,t}$ in the following way:
\begin{equation}\label{eq:unidd_wt}
p_u(q,t) = \left\{ \begin{array}{ll}
p_{0,t} & \textrm{if } q=0\\
p_{k=q/(\mu-1),t} & \textrm{if } \mu-1 \textrm{ divides } q\\
 0 & \textrm{otherwise} \\
\end{array}
\right\}
\end{equation}

\begin{figure}
\centering\resizebox{\columnwidth}{!}{\rotatebox{0}{\includegraphics{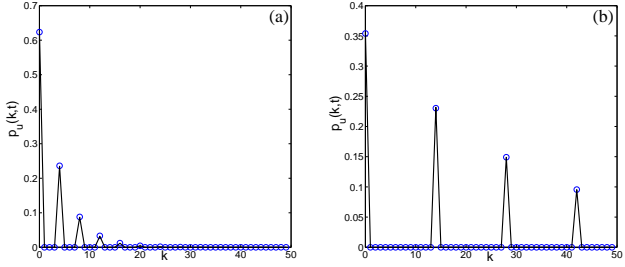}
}}
\caption{Comparison between stochastic simulations for the one-mode
projection (circles) and Eq. (\ref{eq:unidd_wt}) (solid black
curve).
In both the figures $N=500$ and $\gamma=1$. Circles correspond to
averages over $1000$ simulations. In (a) $\mu=5$ while in (b)
$\mu=15$. } \label{fig:unip_multiple}
\end{figure}

Fig. \ref{fig:unip_multiple} shows a comparison between stochastic
simulations (circles) and Eq. (\ref{eq:unidd_wt}) (solid black
curve). Notice that this mapping simply implies that
$p_u(q=0,t)=p_{0,t}$, $p_u(q=\mu-1,t)=p_{1,t}$,
$p_u(q=2(\mu-1),t)=p_{2,t}$, ..., $p_u(q=j(\mu-1),t)=p_{j,t}$.
The same result can be derived by using the generating function
based technique described in Eq. 70 of~\cite{arbit_dd:01}. It is
worth noticing that $q$ is the weighted degree of a node (i.e., the
sum of the weights of all the edges incident on a node), and
therefore, does not give any information about the number of
distinct neighbors a node has.

\subsubsection{Thresholded degree-distribution for parallel attachment}

Weighted graphs, such as the one-mode projections of \abin s, can be
converted to corresponding unweighted version by the process of {\em
thresholding}. A thresholded one-mode projection graph (thresholded
$G_U$) is constructed by replacing every weighted edge in $G_U$ by a
single edge iff the weight of that edge exceeds the threshold value
$\tau$;  otherwise, the edge is deleted. Thresholded degree
distributions are more popular in the complex network literature,
than their weighted counterparts (see, for
example,~\cite{ferrer:04syntax,newman:04}). We shall denote the
thresholded degree distribution at threshold $\tau$ as
$p_u(q,t;\tau)$.

Let us start by considering two nodes $u$ and $u'$ in $U$ with
degrees $k_u$ and $k_{u'}$, respectively. We now try to derive an
expression for the probability $p(k_u,k_{u'}, m)$ that there are
exactly $m$ nodes in $V$ that are linked simultaneously to both $u$
and $u'$. In other words, $p(k_u,k_{u'}, m)$ is the probability that
the number of edges running between $u$ and $u'$ is $m$, given that
the degrees of the nodes are $k_u$ and $k_{u'}$. Let us assume that the
$\mu$ nodes that each node $v \in V$ is connected to, are all
distinct. By the definition of the growth model for \abin s, the
event of $u$ being connected to a node $v$ is independent of $u'$
being connected to the same node. Therefore, the probability that a
randomly chosen node $v \in V$ is connected to $u$ is $k_u/t$ and
the probability that it is connected to $u'$ is $k_{u'}/t$. Recall
that $t$ refers to the number of nodes in $V.$
Thus, the probability that $v$ is connected to both $u$ and $u'$ is
$k_u{k_u'}/t^2$. Therefore, the probability that $u$ and $u'$ share $m$ nodes in $V$ takes the form:
\begin{equation}\label{eq:nodal_wt_dist}
p(k_u,k_{u'},m) = \left( \begin{array}{c} t \\m\end{array}\right)
\left(\frac{k_u k_{u'}}{t^2}\right)^m\left(1-\frac{k_u
k_{u'}}{t^2}\right)^{t-m} \
\end{equation}
From Eq.~(\ref{eq:nodal_wt_dist}), the probability for $u$ and $u'$
of sharing an edge in thresholded $G_U$ is easily computed as:
\begin{eqnarray}
p(k_u,k_{u'}; m > \tau) = \sum_{m=\tau+1}^{t} p(k_u,k_{u'},m) \
\end{eqnarray}
Consequently, in the thresholded $G_U$, the expected degree $D$ of a
node $u$  whose  degree is $k$ in the \abin\ is given by:
\begin{equation} \label{eq:expression_dd}
D(k,\tau) = N\sum_{i=1}^t p_{i,t}\,p(k,i;m > \tau) \
\end{equation}
Notice that then $p_{k,t}$ can be interpreted as the probability of
finding a randomly chosen node with degree $D(k,\tau)$ in the
thresholded one-mode projection.
Thus, the degree distribution of the thresholded $G_U$ is computed
as:
\begin{eqnarray} \label{eq:dd_thresholded_om}
p_u(q,t;\tau) = \sum_{q = \lfloor D(k,\tau) \rfloor} p_k \
\end{eqnarray}
where the function $\lfloor a \rfloor$ returns the largest integer
smaller than $a$.
%

\begin{figure}
\centering\resizebox{\columnwidth}{!}{\rotatebox{0}{\includegraphics{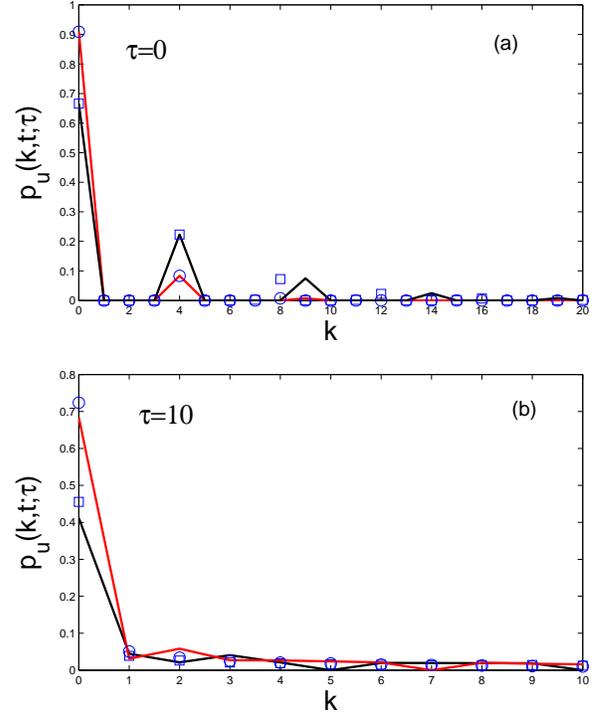}}}
%
\caption{Comparison between stochastic simulations for the one-mode
projection at different times (symbols) and Eq.
(\ref{eq:dd_thresholded_om}) (solid curve). In (a) $\tau=0$,
$N=1000$, $\mu=5$, $\gamma=1$. The circles and the red curve
correspond to $t=20$, while the squares and the black curve to
$t=100$. In (b) $\tau=10$, $N=100$, $\mu=20$, $\gamma=1.5$. The
circles and the red curve correspond to $t=50$, while the squares
and the black curve to $t=100$. } \label{fig:unip_thresholded}
\end{figure}

Fig. \ref{fig:unip_thresholded} shows a comparison between  Eq.
(\ref{eq:dd_thresholded_om}) (solid curves) and stochastic
simulations (symbols) for the one-mode projection at different times.
The implementation of  Eq. (\ref{eq:dd_thresholded_om}) was done by
summing over the $p_{k,t}$ obtained from the stochastic simulations
of the corresponding \abin\ according to $q = \lfloor D(k,\tau) \rfloor$, as indicated by Eq. (\ref{eq:expression_dd}).

\subsubsection{One-mode kernel}

Until now we have been describing growth models for the \abin s. The
unipartite network $G_U$ is obtained by projecting the \abin\ onto
the set of nodes $U$. We shall now attempt to derive a kernel for
the growth of the network $G_U$, whereby we can construct $G_U$
directly without constructing the underlying \abin . Consider a node
$v_t \in V$ that has been introduced in the \abin\ in the $t^{th}$
step. There are $\mu$ nodes in $U$ to which $v_t$ gets connected.
Let us assume that $v_t$ is connected to no node in $U$ more than
once. This fact is true in the ``parallel attachment without
replacement" model that will be described in greater details in
Sec.~\ref{sec:planet}. However, as discussed earlier, if $\mu \ll N$
and $\gamma$ is small, it is quite reasonable to make this
assumption even in the case of ``parallel attachment with
replacement" model.

Introducing $v_t$ in the \abin\ is equivalent to
introducing a clique (complete graph) of size $\mu$ in $G_U$. This
is because all the nodes that are connected to $v_t$ in the \abin\
are connected to each other in $G_U$ by virtue of sharing a common
neighbor $v_t$. Note that this does not prohibit these $\mu$ nodes
from having previous connections. 
The growth process is such that 
 multiple edges, or equivalently edge weights 
between two nodes larger then 1 can occur.

Let us denote the degree of a node $u_i$ in (the non-thresholded)
$G_U$ after $t$ steps as $q_{i,t}$. As discussed in the previous
subsection, $q_{i,t} = (\mu-1)k_{i,t}$, where $k_{i,t}$ is the
degree of $u_i$ in the corresponding \abin\ after $t$ steps.
Noticing the fact that in the \abin\ the $\mu$ nodes are chosen
independently of each other solely based on the attachment kernel,
we can define a kernel for selecting a set of $\mu$ nodes in $G_U$
as follows.
\begin{equation}\label{eq:onemode_kernel}
\widetilde{A}(q_{a,t},q_{b,t}, \dots) = \prod_{j=a,b,\dots}
\widetilde{A}(q_{j,t}/(\mu-1)) \
\end{equation}
where $a,b,\dots$ denotes a randomly chosen set of $\mu$ nodes in
$G_U$.
Substituting the expression provided in
Eq.~(\ref{eq:kernel_attachment}) for the preferential attachment
based kernel we obtain:
\begin{equation}\label{eq:onemode_lpa}
\widetilde{A}(q_{a,t},q_{b,t}, \dots) = \prod_{j=a,b,\dots}
\frac{\gamma/(\mu-1) q_{j,t} + 1}{\sum_{i=1}^{N} (\gamma/(\mu-1)
q_{i,t} + 1) } \
\end{equation}
Below we summarize the growth model for the one-mode projection of
the \abin\
\begin{itemize}
\item Select a set of $\mu$ nodes $a,b,\dots$ with the probability  $\widetilde{A}(q_{a,t-1},q_{b,t-1}, \dots)$ as 
described by Eq. (\ref{eq:onemode_lpa}).
\item Introduce edges between every pair of the chosen set $a,b,\dots$.
\item Advance time by a unit and repeat the process.
\end{itemize}
We assume an  initial condition $q_i=0$ for all $i$. Alternatively,
but also equivalently, the above growth model can be described as
choosing $\mu$ nodes independently, each with probability
$\widetilde{A}(q_{i,t}/(\mu-1))$ and then adding edges between them.

\begin{figure}
\centering \centering
\resizebox{\columnwidth}{!}{\rotatebox{0}{\includegraphics{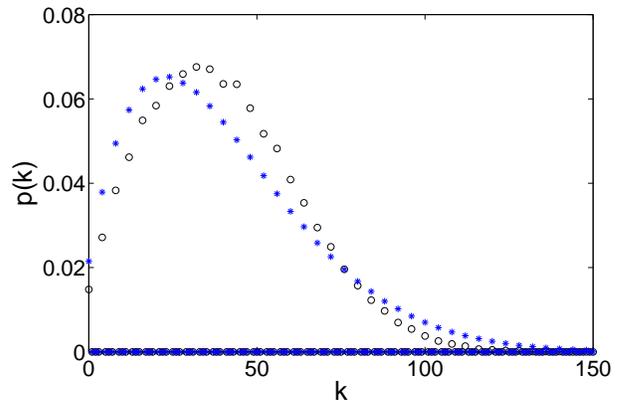}}}
\caption{Comparison between the degree distribution obtained from
the stochastic simulation of Eq. (\ref{eq:onemode_lpa}), averaged
over 1000 runs, and the one-mode projection of the \abin\ obtained using Eq.
(\ref{eq:unidd_wt}), averaged over 100000 runs, with $N=50$, $\mu=5$,
$\gamma=0.5$ at $t=100$. Circles correspond to the one-mode kernel
degree distribution, i.e., Eq. (\ref{eq:onemode_lpa}), while the stars are the one-mode projection of the \abin.}\label{fig:omkernel}
\end{figure}

Fig.~\ref{fig:omkernel} plots the degree distribution obtained from
the one-mode kernel and the degree distribution of the one-mode
projection of the $u_i$ nodes of the \abin\ built with the same
parameters. We can see that the one-mode kernel gives quite similar
degree distribution as one-mode projection of the bipartite network.
The primary observation from this analysis is that the kernel of the
unipartite growth model has the same form as that of the bipartite
growth model, with a scaling of the parameter $\gamma$ by a factor
of $1/(\mu-1)$ in the former. This implies that as $\mu$ increases,
the extent of degree-based preference decreases in the one-mode
projection. The analysis, nevertheless is valid only for the
``without replacement" model and holds approximately for the ``with
replacement" for $\mu \ll N$.

\section{Real world \abin}\label{sec:realwork}

\subsection{CoGNet: the codon-gene network}\label{sec:cogen}

As complete genomes of more and more organisms are sequenced,
phylogenetic trees reconstructed from genomic data become increasingly detailed. 
Codon usage patterns in different genomes can provide insight into phylogenetic
relations. However, except for some earlier work~\cite{Sharp88},
studies on the codon usage have not received much attention. One of
the main research issues in this context is to understand the
influence of randomness in the growth pattern of genome sequences in
the context of biological evolution. A well known random process in
evolutionary biology is {\em random mutation} in a gene sequence. A
gene sequence is a string defined over four symbols (A, G, T, and C)
that represent the nucleotides. A {\em codon} is a triplet of
adjacent nucleotides (eg. AGT, CTA) and codes for a specific amino acid.
There are only 64 codons. Codon usage in genome sequences varies
between different phylogenetic groups.

\subsubsection{Definition and construction}

\begin{table*}
\centering \caption{List of organisms along with their probable
origin time (in Million Years Ago current time) and codon and
gene counts} \label{orglist}
 \centering
\begin{tabular}{|c|c|c|c|c|} \hline
   Organism's Name & Description & Origin time (MYA) & Gene count & Codon count \\ \hline \hline
  \emph{Myxococcus xanthus}  & Gram-negative rod-shaped bacterium & 3200 & 7421 & 2822743 \\ \hline
  \emph{Dictyostelium discoideum}  & Soil-living amoeba & 2100 & 3369 & 1962284 \\ \hline
  \emph{Plasmodium falciparum}  & Protozoan parasite & 542 & 4098 & 3032432 \\ \hline
  \emph{Saccharomyces cerevisiae}  & Single-celled fungi & 488& 14374 & 6511964 \\ \hline
  \emph{Xenopus laevis} & Amphibian, african clawed frog & 416 & 12199 & 5313335 \\ \hline
  \emph{Drosophila melanogaster}  & Two-winged insect, fruit fly & 270 & 40721 & 21393288 \\ \hline
  \emph{Danio rerio}  &  Tropical fish, zebrafish & 145 & 19062 & 8042248 \\ \hline
  \emph{Homo sapiens}  & Bipedal primates, Human & 2 & 89533 & 38691091 \\ \hline
\end{tabular}
\end{table*}

We refer to the  network of codons and genes as CoGNet and represent
it as an \abin\, where $V$ is the set of {\em genes}, i.e., genome
of the organisms, and $U$ is the set of nodes labeled by the codons.
There is an edge $(u,v) \in E$ that run between $V$ and $U$ if and
only if the codon {\em u} occurs in the gene {\em v}.
Fig.~\ref{fig:dcslife} illustrates the structure of CoGNet.

We have analyzed 8 organisms belonging to widely different
phylogenetic groups. These organisms have been
extensively studied in biology and genetics~\cite{Hedges02} and, for
our purpose importantly, their genomes have been fully sequenced. In
Table~\ref{orglist} we list these organisms along with a short
description and the number of genes (i.e., the cardinality of set
$V$) and codons (i.e., the cardinality of set $U$). The data have
been obtained from the Codon Usage Database~\cite{Nakamura00,
codonsite}. The usage of a particular codon in an organism's genome
sequence can be as high as one million. In other words, the degree
of the nodes in $U$ can be arbitrarily large. This, together with the
fact that there are only 64 nodes in $U$, presents us with the
non-trivial task of estimating the probability
distribution $p_k$, having a very large event space (between 0 and
few millions), from very few observations (only 64).

A possible strategy to cope with this situation is through {\em
binning} of the event space. For example, if we use a bin size of
$10^4$, then degree $1$ to degree $10^4$ is compressed to a single
bin which we label as $1$, the next $10^4$ degrees are mapped into
the bin $2$, and so on. Thus, if for a particular organism the codon
count is $m$, then theoretically, the maximum degree of a codon node
can be $m$, which in turn implies that with a bin size of $10^4$,
there will be $m/10^4$ bins (or possible events) in which the 64
data points will be distributed. If all organisms are analyzed using
the same bin size, depending on the length of the organism's genome,
i.e., the codon count $m$, one obtains different number of bins.
Alternatively, the bin size can be set for each organism in such a
way that the resulting number of bins remains the same for all
organisms. Thus, if we wish to have $b$ bins for all organisms, the
bin size for a particular organism will be $m/b$. Here we analyze
the data using both the methods: fixed bin size and fixed number of
bins.

Apart from {\em binning}, another way to cope with the problem of
data sparseness is to compute the cumulative degree distribution
$P_{k,t}$ rather than the standard degree distribution $p_{k,t}$.
$P_{k,t}$ is defined as the probability that a randomly chosen node
has a degree less than or equal to $k$. Thus,
\begin{equation}\label{eq:def_cumulative}
P_{k,t} = \sum_{i=0}^{k} p_{i,t} \,.
\end{equation}
The cumulative distribution is more robust to noise present in the
observed data points, but at the same time it contains all the
information present in $p_{k,t}$~\cite{newman03}. Note that
even though it is a standard practice in statistics to define
cumulative distribution as stated in Eq.(\ref{eq:def_cumulative}),
in complex network literature it is defined as the probability that
a randomly chosen node has degree ``greater than or equal to" $k$.

\begin{figure}
\centering\resizebox{\columnwidth}{!}{\rotatebox{0}{\includegraphics{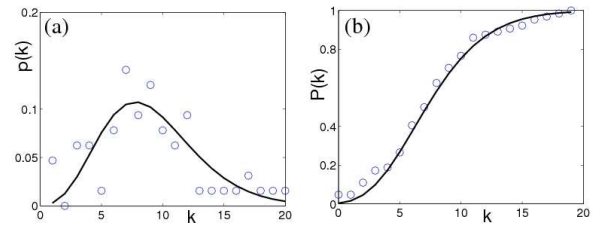}}}
\caption{Degree distribution of the codon nodes for \emph{Xenopus
leavis}. In (a) a comparison between the real data (symbols) and the
theoretical $p_{k,t}$ obtained using Eq. (\ref{eq:solutionparallel})
(black solid curve) is shown. The cumulative distribution of the
real data (symbols) and the theory (black solid curve) is shown in
(b). } \label{fig:fitting}
\end{figure}

Fig.\ref{fig:fitting}(a) shows a comparison between the empirical
degree distribution for Xenopus leavis (symbols) and the
corresponding theoretical distribution predicted by
Eq.(\ref{eq:solutionparallel}) at a $\gamma$ for which the squared
error between the two distributions is minimum.
Fig.\ref{fig:fitting}(b) presents the same data, but in terms of the
cumulative distribution. 

\begin{table*}
\centering \caption{The values of $\gamma$ that yield best fit for
the degree distribution under the two different {\em binning}
strategies.}\label{org table}
 \centering
\begin{tabular}{|c|c|c|} \hline
   Organism's Name & Best $\gamma$ (fixed bin size) & Best $\gamma$ (fixed bin count)  \\ \hline \hline
  \emph{Myxococcus xanthus}  & 2.35 & 2.1  \\ \hline
  \emph{Dictyostelium discoideum}  & 2.38 & 2.57 \\ \hline
  \emph{Plasmodium falciparum}  & 1.36 & 1.81 \\ \hline \hline
  \emph{Saccharomyces cerevisiae}  & 0.35 & 0.34 \\ \hline
  \emph{Xenopus laevis}  & 0.11 & 0.11 \\ \hline
  \emph{Drosophila melanogaster}  & 0.28 & 0.2 \\ \hline
  \emph{Danio rerio}  & 0.14 & 0.1 \\ \hline
  \emph{Homo sapiens} & 0.20 & 0.09 \\ \hline
\end{tabular}
\end{table*}

\subsubsection{Growth model}

A particular gene does not acquire all its constituent
codons at a single time instance but evolves from an ancestral gene
through the process of mutation (addition, deletion  or substitution
of codons)~\cite{kunkel00}. Therefore, we choose to apply the ``sequential
attachment" based growth model for synthesis of CoGNet. 
This means that we model the CoGNet growth through equations (\ref{eq:markovchain}) and (\ref{eq:solutionparallel}).

For all the CoGNets, the value of $N$ is $64$, $\mu$ is 1 and $t$
corresponds to the number of codons that appears in the genome of the organism. In our model, we have a single
fitting parameter, $\gamma$; The value of $\gamma$ is chosen such that
the difference or error between the distributions obtained from the empirical data
and the synthesized CoGNet is minimized. The error, $E$, is defined as
follows.
\begin{equation} \label{eq:def_error}
E = \sum_{k=0}^{\infty}( p_{k,t}(\gamma) - p^*_{k,t} )^2  \,,
\end{equation}
where $p^*_{k,t}$ represents the empirical distribution.

\begin{figure*}
\centering\resizebox{\linewidth}{!}{\rotatebox{0}{\includegraphics{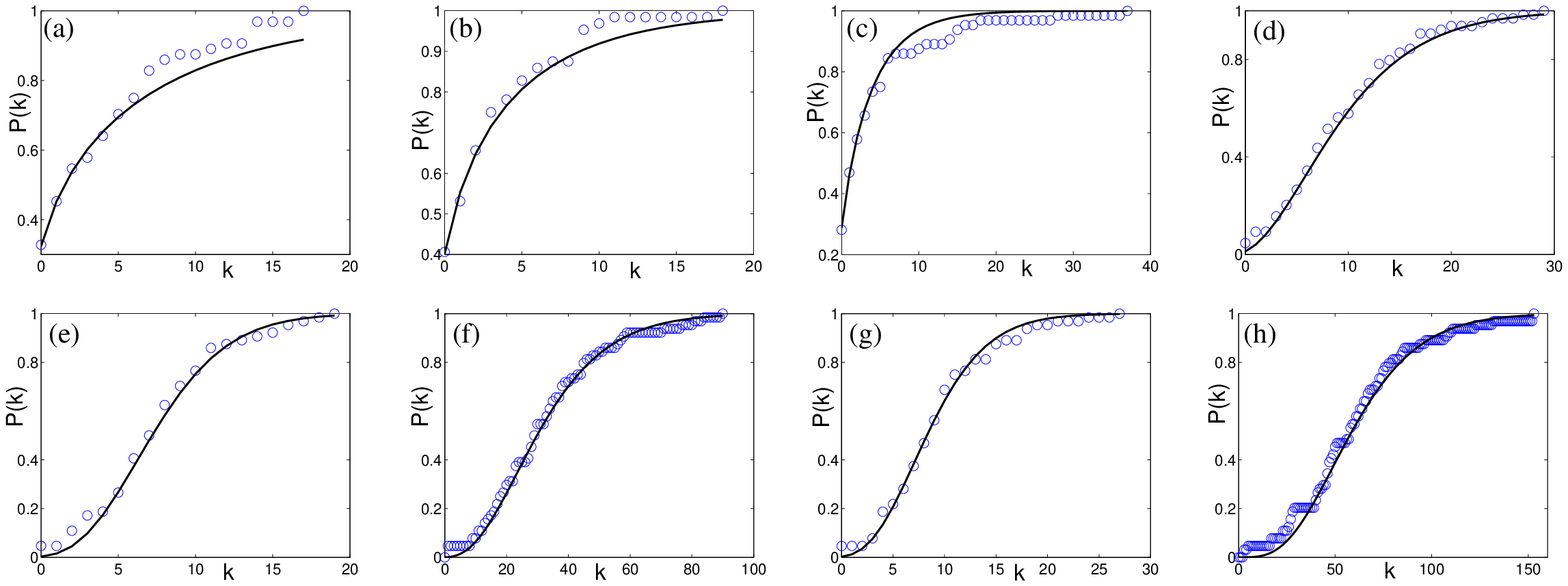}}}
%
%
%
\caption{Cumulative degree distributions for the empirical data
(symbols) and their corresponding theoretical best $\gamma$-fits through Eqs. (\ref{eq:markovchain}) and (\ref{eq:solutionparallel}) (solid
curve) for the organisms.  (a) \emph{Myxococcus xanthus}, (b)
\emph{Dictyostelium discoideum}, (c) \emph{Plasmodium falciparum},
(d) \emph{Saccharomyces cerevisiae}, (e) \emph{Xenopus laevis}, (f)
\emph{Drosophila melanogaster}, (g) \emph{Danio rerio}, and (h)
\emph{Homo sapiens}.} \label{fig:dist_organism}
\end{figure*}

Fig.~\ref{fig:dist_organism} shows the cumulative real data and
corresponding theoretical distributions of the eight organisms listed
in Table~\ref{orglist}. Table.~\ref{org table} lists the values of
$\gamma$ for two different methods of binning: fixed bin count (bin
count = 20) fixed bin size (bin size = $10^4$).

It can be observed that the values of $\gamma$ get polarized into
two distinct groups. The value of $\gamma$ for {\em binning} with
fixed bin size is much higher for three organisms (between 1.36 and
2.38), that are simple and primitive, than the rest (between 0.11
and 0.35) which are more complex and came into existence at a later
stage of evolution. In order to test whether bin size might influence the value of
$\gamma$, the experiments were repeated with various bin sizes. The
analysis reveals that the polarization of the organisms into two
classes based on the value of $\gamma$ is almost independent of the bin size.

We conclude that at least at the level of codon usage in {\it Myxococcus xanthus}, {\it Dictyostelium discoideum}, and {\it Plasmodium falciparum} the degree of randomness during codon selection is much lower than in  {\it Saccharomyces cerevisiae}, {\it Xenopus laevis}, {\it Drosophila melanogaster}, {\it Danio rerio}, and {\it Homo sapiens}. 
These findings are probably correlated to the origin time and the evolutionary processes that shaped the usage of codons as follows.  
Let us think of evolution as the product of ``copy-paste" operations. 
In this way, new genes emerge as result of defectous copy-paste operations where the ancestral genes that are being copied are altered by addition, deletion or substitution of codons. 
Thus, copy-paste operations without defects lead to a high degree of ``preferential attachment", while mutations/deffects increase the degree of randomness. 
In consequence, we expect newly born species/organisms to exhibit a higher degree of randomness than their ancestor, given the greater number of mutations experienced by the newly formed organisms. 
The value of $\gamma$ in Table.~\ref{org table} reflects this fact, and suggests that knowledge at the level of codon usage (i.e., $\gamma$) can be used as a criterion to classify organisms.

\subsection{PlaNet: the phoneme-language network}\label{sec:planet}

In this section, we attempt to explain the self-organization of the
consonant inventories through \abin\, where the consonants make up
the basic units and languages are thought as discrete combinations
of them. In fact, the most basic units of human languages are the
speech sounds. The repertoire of sounds that make up the sound
inventory of a language are not chosen arbitrarily. Indeed, the
inventories show exceptionally regular patterns across the languages
of the world, which is arguably an outcome of the self-organization
that goes on in shaping their structures~\cite{Oudeyer:06}.
In order to explain this self-organizing behavior of the sound
inventories, various functional principles have been proposed such
as {\em ease of articulation}~\cite{Lindblom:88,Boer:00}, {\em
maximal perceptual contrast}~\cite{Lindblom:88} and {\em
learnability}~\cite{Boer:00}.
The structure of vowel inventories has been successfully explained
through the principle of maximal perceptual
contrast~\cite{Lindblom:88,Boer:00}.
Although there have been some linguistically motivated work
investigating the structure of the consonant inventories, most of
them are limited to certain specific properties rather than
providing a holistic explanation of the underlying principle of its
organization.

\subsubsection{Definition and construction}

\begin{figure}[!t]
\begin{center}
\centering\resizebox{\columnwidth}{!}{\rotatebox{0}{\includegraphics{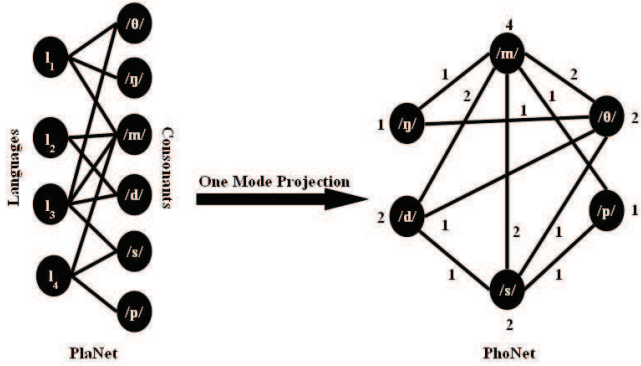}}}
\caption{Illustration of the nodes and edges of PlaNet and PhoNet.}
\label{fig:planet}
\end{center}
\end{figure}

A first study of the consonant-language network as an \abin\ can be
found in~\cite{choudhury:06}.
Here we follow the same definitions given in~\cite{choudhury:06} and
refer to the consonant-language \abin\ as PlaNet or Phoneme-Language
Network.
$U$ is the universal set of consonants and $V$ is the set of
languages of the world.
There is an edge $(u,v) \in E$ iff the consonant {\em u} occurs in
the sound inventory of the language {\em v}.
On the other hand, the one-mode projection of PlaNet onto the
consonant nodes is called PhoNet.
Fig.~\ref{fig:planet} illustrates the structures of PlaNet and
PhoNet. Note that PlaNet is an unweighted bipartite graph, whereas
PhoNet has been represented as a weighted graph.

Many typological studies~\cite{Hinskens:03,Ladefoged:96,Lindblom:88} of segmental inventories have been carried out in 
the past on the UCLA Phonological Segment Inventory Database (UPSID)~\cite{Maddieson:84}. UPSID records the sound 
inventories of 317 languages covering all the major language families of the world. In this work, we have used UPSID 
consisting of
these 317 languages and 541 consonants found across them, for
constructing PlaNet. Consequently, there are 317 elements (nodes) in
the set $V$ and 541 elements (nodes) in the set $U$. The number of
elements (edges) in the set $E$ as computed from PlaNet and PhoNet
are 7022 and 30412 respectively.
We selected UPSID mainly due to two reasons -- (a) it is the largest
database of this type that is currently available and, (b) it has
been constructed by selecting one language each from moderately
distant language families, which ensures a considerable degree of 
``genetic" balance.

\subsubsection{Topological properties}

\begin{figure}
\begin{center}
\centering\resizebox{\columnwidth}{!}{\rotatebox{0}{\includegraphics{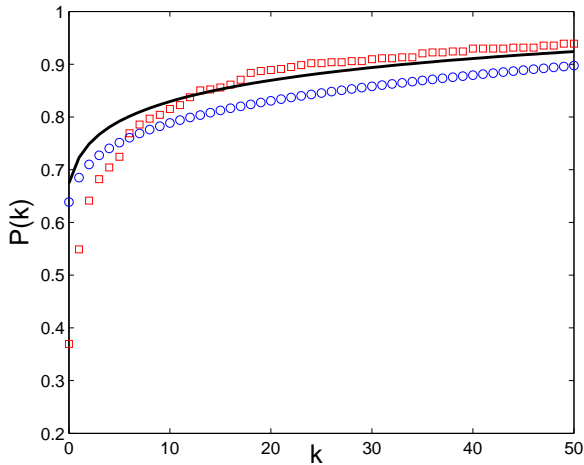}}}
\caption{Cumulative degree distribution of $U$, i.e., the consonant
nodes. Squares correspond to the empirical data, and circles to simulations performed with ``parallel attachment without
replacement" with $\gamma=14$  (PlaNet$_{sim}$).The solid line corresponds
to the theoretical solution for ``parallel attachment with
replacement" (PlaNet$_{theo}$) obtained through integration of Eq.
(\ref{eq:parallel_full}) with $\gamma=14$.} \label{fig:dd_cons}
\end{center}
\end{figure}

Fig.~\ref{fig:dd_cons} illustrates the (cumulative) degree distribution of $U$.
%
Since the degree of a language node is nothing but the
size of the consonant inventory, we take as $\mu$, i.e., the degree of each $V$ node, the average number of consonants in human languages which is $22$. 
Recall that in the theory for
\abin\ the degree of each node in $V$ has been assumed to be a
constant $\mu$.

\subsubsection{Growth models}
In order to obtain a theoretical description of the degree
distribution of the consonant nodes in PlaNet (and later on PhoNet),
 we employ the \abin\ growth model described in Sec.~\ref{sec:growthmodel}.
We assume that all the language nodes have a degree $\mu = 22$.
Clearly, $N= 541$ is the total number of consonant nodes and $t =
317$ is the total number of languages.
Thus, $\gamma$ is the only free parameter in the model.
Notice that, by definition, in PlaNet a consonant can occur only
once in a language inventory.
Therefore, unlike the case of CoGNet, PlaNet is an \abin\ that has
been constructed using a ``parallel attachment without replacement"
scheme.
However, we expect the theory developed in
Sec.~\ref{sec:growthmodel}, corresponding to ``parallel attachment
with replacement", to be a fairly good approximation for the degree distribution of PlaNet. We
shall refer this theoretical model of PlaNet as PlaNet$_{theo}$.
In order to estimate the free parameter $\gamma$, the best fit was
obtained with $\gamma=14$ (see Fig. \ref{fig:dd_cons}). Since
$1\le\gamma\le N/\mu= 24.6$, based on our theoretical analysis we
can conclude that the attachments are largely preferential in nature
and the degrees follow a beta distribution with a single mode at
$k=1$.

To study the effect of the ``parallel attachment without
replacement" scheme, we carry out stochastic simulations with such a
growth model described below. Suppose that a language node $v_i$
(with degree 22) is added to the system and that $j < 22$ edges of
the incoming node have already been attached to $u_1,\,u_2,...,u_j$
distinct consonant nodes.
Then, the $(j+1)$th edge is attached to a consonant node based on
the same preferential attachment kernel (see Eq.
\ref{eq:kernel_attachment}), but applied on the reduced set $U -
\{u_1, u_2, \dots, u_j\}$, i.e., the previously selected
$u_1,\,u_2,...,u_j$ consonant nodes cannot participate in the
selection process of the $(j+1)$th edge of $v_i$. This ensures that
a consonant node is never chosen twice. We shall refer to the degree
distributions of the consonant nodes obtained in this way as
PlaNet$_{sim}$. The degree distribution of PlaNet$_{sim}$ has the
best match with the degree distribution of the real PlaNet when
$\gamma = 14$.

We have calculated the error for the aforementioned stochastic
simulation model ($E_{sim}$) as well as the theory of
Sec.~\ref{sec:growthmodel}, corresponding to ``parallel attachment
with replacement" ($E_{theo}$). The error has been computed using
Eq. (\ref{eq:def_error}) where $p^*_{k,t}$ stands for the degree
distribution of the real PlaNet. It is found that $E_{sim}=0.0972$
and $E_{theo}=0.1170$. Since the simulation using the ``parallel
attachment without replacement'' scheme describes the structure of
consonant inventories better, the error in this case is smaller than
that for ``parallel attachment with replacement".

\subsubsection{One-mode projection: PhoNet}
\begin{figure}
\begin{center}
\centering\resizebox{\columnwidth}{!}{\rotatebox{0}{\includegraphics{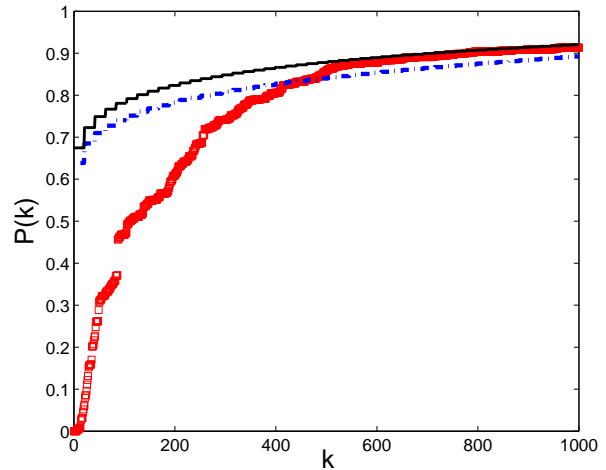}}}
\caption{Cumulative degree distribution of the one-mode projection
of PlaNet (PhoNet). Squares correspond to the empirical data
(Real PhoNet), dash-dotted line to simulations of one-mode
projection model with ``attachment without replacement" 
using kernel Eq.~(\ref{eq:kernel_attachment}) (PhoNet$_{sim}$). 
The solid curve shows the theoretical degree distribution with
the ``attachment with replacement" scheme using
Eq.~(\ref{eq:mu_recursion}) (PhoNet$_{theo}$).}
\label{fig:phonet_sym}
\end{center}
\end{figure}

Interestingly, when we reconstruct the one-mode projection from
either the theory  using the ``attachment with replacement" scheme
(PhoNet$_{theo}$) or stochastic simulation considering the
``attachment without replacement" model (PhoNet$_{sim}$), we cannot
match the empirical data.
Fig.~\ref{fig:phonet_sym} shows the cumulative degree distributions
of   PhoNet$_{sim}$, PhoNet$_{theo}$ and real PhoNet. We have
calculated the error of  PhoNet$_{sim}$ and PhoNet$_{theo}$ with
respect to the real PhoNet using the Eq. (\ref{eq:def_error}) and
refer them as ($E_{sim}$) and ($E_{theo}$) respectively. Experiments
reveal that $E_{sim}$ = 0.1230 and $E_{theo}$ = 0.1438.
The results show a larger quantitative difference between the curves
compared to that between their bipartite counterparts. It indicates
that the one-mode projection has a more complex structure than that
could have emerged from a simple preferential attachment based
kernel.

Anyway, we observe that preferential attachment can explain the
occurrence distribution of the consonants over languages to a good
extent. One possible way to explain this observation would be that a
consonant, which is prevalent among the speakers of a given
linguistic generation, tends to be more prevalent in the subsequent
generations with a very little randomness involved in this whole
process. It is this micro-level dynamics that manifests itself as
preferential attachment in PlaNet. However, the fact that the
co-occurrence distribution of the consonants, i.e., the degree
distribution of PhoNet, is not explained by the growth model implies
that there are other organizing principles absent in our current
model that are involved in shaping the structure of the consonant
inventories.


\section{Discussion and Conclusion}\label{sec:disc}

In the preceding sections, we have presented growth models for
discrete combinatorial systems in the framework of a  special class
of networks -- \abin s. To summarize some of our important
contributions, we have
\begin{itemize}
\item proposed growth models for \abin s, which are based on preferential
attachment coupled with a tunable randomness component,
\item extended the mathematical analysis presented in~\cite{us_epl} and 
derived the exact expression for the degree distribution in case of
parallel attachment,
\item analytically derived the degree distribution of the one-mode
projection,
\item and presented case studies for two well-known DCSs from the domain
of biology and language and, thereby, we have validated
our analytical findings against the empirical data.
\end{itemize}

It is worthwhile to mention here that there have been certain
alternative perspectives of viewing the DCS problem presented here. One
of the most celebrated among these is the ``P{\'o}lya's Urn" model
(see~\cite{johnson:77}). 
In this classical model there is an urn
initially containing $r$ red and $b$ blue balls. One ball is chosen
randomly from the urn. The ball is then put back into the urn
together with another new ball (presumably from a collection stored
elsewhere) of the same color. Hence, the number of total balls in
the urn grows. Generalizations of this classical model have been
proposed and solved by Chung et al. in~\cite{chung:03}. In this
model, the authors assume that there are finitely many urns each
containing one ball and the additional balls arrive one at a time.
With each new incoming ball, a new urn is created with a probability
$p$ and the ball is placed in this newly created urn. With
probability 1-$p$ the ball is placed in an existing urn, where the
probability that an urn, currently containing $m$ balls, is chosen
for placing the new ball is proportional to $m^{\nu}$. Note that,
for $p = 0$, the number of urns is fixed and finite and the model
resembles the one we proposed here; however, in this case the
tunable randomness component $\gamma$, which is the most important
parameter of our model, is absent. From the analysis of this model
the authors find that for $\nu < 1$, the balls in all the urns grow
at roughly the same rate. For  $\nu > 1$, one urn dominates, i.e.,
the probability that any new ball goes into that urn is equal to 1.
For  $\nu = 1$, the fraction of balls going into each urn converges,
though the limit is uniformly distributed in a certain simplex
(see~\cite{chung:03} for proofs). Here we have derived the exact
analytical form for the probability distribution of the number of
urns with a specified number of balls ($k$) after the addition of $t$
balls. Moreover the proposed model takes into account a tunable
randomness parameter, as well as the case where more than one ball
are placed into the urns simultaneously (parallel attachment).

Another important issue that needs a mention is that although the reported results are strictly 
valid for a set of basic units fixed in time, we argue here this condition can be relaxed. 
We can find some real systems where the set of basic units also grow,
however, at a far slower rate than the set of their discrete
combinations. Under this condition we can expect the reported results to approximately hold as long as the growth rate of the basic units is slow enough.

Finally, as this study reveals, there are certain limitations of the
growth models proposed here. For instance, it has been shown through
simulations that the degree distribution of the consonant nodes in
PlaNet is better explained by having a superlinear kernel as opposed
to a linear kernel introduced here~\cite{Mukherjee07soundInventory}.
An analytical treatment of such a superlinear kernel should be an
interesting topic for future research. There are also some
limitations in the study of CoGNet. Selection of correct binning
policy to construct the CoGNet is a challenging job. Modeling the
CoGNet with parallel attachment where $\mu$ is the average number of
codons present in the genes is a direct extension of the current
work. As a first step, we here classified the eight organisms into two sets
and we believe that our new method can further contribute to the 
reconstruction of phylogenetic relations. Our approach may be especially 
useful for the analysis of such genome sequences which are so far only 
available in fragments either due to fragmentary sampling of the biological 
material or to un-finished sequencing efforts. 

\acknowledgments
This work was partially financed by the Indo-German collaboration project DST-BMBT through grant ``Developing robust and efficient services for open
source Internet telephony over peer to peer network".
N.G., A.N.M. and A.M. acknowledge the hospitality of TU-Dresden. 
F.P. acknowledges the hospitality of IIT-Kharagpur and funding through grant ANR BioSys (Morphoscale).


\end{document}